\documentclass[conference]{IEEEtran}
\IEEEoverridecommandlockouts
\usepackage{etoolbox}
\usepackage{tikz}

\usepackage{listings}

\usepackage{graphicx} 
\usepackage{enumitem}
\usepackage{circledsteps}
\usepackage{subcaption}
\usepackage{multirow}
\usepackage{hhline}
\usepackage{lipsum}
\usepackage{rotating}
\usepackage{algorithm}
\usepackage{algpseudocode}
\usepackage{soul}
\usepackage{cite}
\usepackage{amsmath,amssymb,amsfonts}
\usepackage{graphicx}
\usepackage{textcomp}
\usepackage{xcolor}
\usepackage{tcolorbox}
\usepackage{booktabs}
\newbool{showComments}
\booltrue{showComments}
\ifbool{showComments}{%
\newcommand{\rob}[1]{\textcolor{blue}{\textbf{RM:} #1}}
\newcommand{\sy}[1]{\textcolor{orange}{\textbf{SJ:} #1}}
\newcommand{\deleteifnospace}[1]{{\textcolor{red}{#1}}}

}{
\newcommand{\rob}[1]{}
\newcommand{\sy}[1]{}
\newcommand{\deleteifnospace}[1]{}

}

\def\BibTeX{{\rm B\kern-.05em{\sc i\kern-.025em b}\kern-.08em
    T\kern-.1667em\lower.7ex\hbox{E}\kern-.125emX}}
\begin{document}

\title{Edge-First Language Model Inference: \\Models, Metrics, and Tradeoffs}

\author{\IEEEauthorblockN{SiYoung Jang}
\IEEEauthorblockA{\textit{Nokia Bell Labs} \\
Cambridge, United Kingdom \\
siyoung.jang@nokia-bell-labs.com}
\and
\IEEEauthorblockN{Roberto Morabito}
\IEEEauthorblockA{\textit{EURECOM} \\
Biot, France \\
roberto.morabito@eurecom.fr}
}

\maketitle

\begingroup
\renewcommand\thefootnote{}\footnotetext{\noindent\footnotesize This paper has been accepted for publication and presentation at the 45th IEEE International Conference on Distributed Computing Systems (IEEE ICDCS 2025). The copyright will be transferred to IEEE upon publication in the conference proceedings.}%
\addtocounter{footnote}{-1}
\endgroup

\begin{abstract}
The widespread adoption of Language Models (LMs) across industries is driving interest in deploying these services across the computing continuum, from the cloud to the network edge. This shift aims to reduce costs, lower latency, and improve reliability and privacy. Small Language Models (SLMs), enabled by advances in model compression, are central to this shift, offering a path to on-device inference on resource-constrained edge platforms. This work examines the interplay between edge and cloud deployments, starting from detailed benchmarking of SLM capabilities on single edge devices, and extending to distributed edge clusters. We identify scenarios where edge inference offers comparable performance with lower costs, and others where cloud fallback becomes essential due to limits in scalability or model capacity. Rather than proposing a one-size-fits-all solution, we present platform-level comparisons and design insights for building efficient, adaptive LM inference systems across heterogeneous environments.
\end{abstract}


\vspace{-0.5cm}
\section{Introduction}

In recent years, Large Language Models (LLMs) have demonstrated remarkable capabilities in handling a wide array of language comprehension tasks, including answering complex queries, arithmetic reasoning, contextual understanding, sentiment analysis, text categorization, factual retrieval, and code generation. These models have revolutionized numerous applications by showcasing unprecedented accuracy and contextual awareness in natural language processing (NLP) tasks. However, their computational intensity—driven by self-attention mechanisms and large-scale training requirements—makes them best suited for cloud-hosted deployment, where storage, compute, and scalability are readily available.

Despite their success, cloud-based LLMs come with significant drawbacks, including dependency on consistent network connectivity, latency issues, high operational costs, and privacy concerns. For instance, cloud-based inference introduces non-negligible round trip times which may hinder real-time or mission-critical applications (see Section~\ref{sec:jointinference} for detailed measurements). Additionally, service availability can be impacted by rate limits or server overloads, and user prompts may be logged or used for model training, further amplifying privacy concerns~\cite{openaiuserdatausage}. These limitations are fueling a shift towards edge-centric solutions.

Central to this shift is the growing interest in Small Language Models (SLMs), which, enabled by techniques like quantization, pruning, and distillation, allow language capabilities to be executed on resource-constrained edge devices such as smartphones~\cite{laskaridis2024melting,xu2023llmcad}. By reducing model size or precision~\cite{lin2024awq-quantization, ma2023llmpruner, gu2024llm-kd}, SLMs make on-device inference viable, addressing many of the challenges posed by centralized LLMs, particularly in terms of cost, latency, and data locality. However, SLMs still fall short of matching the performance and generality of their larger counterparts. 
This leads to a fundamental question: 

\begin{quote}
\textit{How can SLMs be leveraged at edge to reduce cloud dependency while maintaining performance, privacy and responsiveness?}
\end{quote}

In this work, we adopt an edge-first perspective and explore how to deploy and evaluate SLMs across a distributed edge environment. We begin by benchmarking the performance, responsiveness, and energy consumption of opensourced SLMs on off-the-shelf edge devices and extend our investigation to a distributed edge cluster. We then assess conditions under which cloud fallback becomes necessary due to limitations in edge scalability. Rather than promoting a one-size-fits-all deployment, we focus on the tradeoffs across model types, hardware platforms, and performance-cost metrics—reflecting the core theme of this paper: \textit{Models, Metrics, and Tradeoffs}.

\section{Background}~\label{sec:jointinference}
 \label{subsec:cloud-llm}
Conventional \textbf{cloud-based LLM} services such as ChatGPT~\cite{openai_chatgpt} offers users with highly accurate response, to demonstrate impressive capabilities in tasks. 
For example, ChatGPT-4o demonstrates impressive benchmark scores across various areas, including general knowledge (MMLU)
, mathematical and logical reasoning (GSM8K)
, coding (HumanEval)
, while also supporting audio and visual multimodal capabilities~\cite{openai_gpt4o_2024}. 
Additionally, cloud-based LLM inference offers significant advantages in terms of accessibility and scalability. 
By hosting language models in the cloud, users can utilize the service via internet connectivity, regardless of their geographical location. 
Moreover, cloud infrastructures offers dynamic resource allocation, automatically adapting compute resources to workload demands without requiring user intervention, simplifying end-user experience.

\begin{figure}[!t]
    \centering
    \includegraphics[width=\columnwidth]{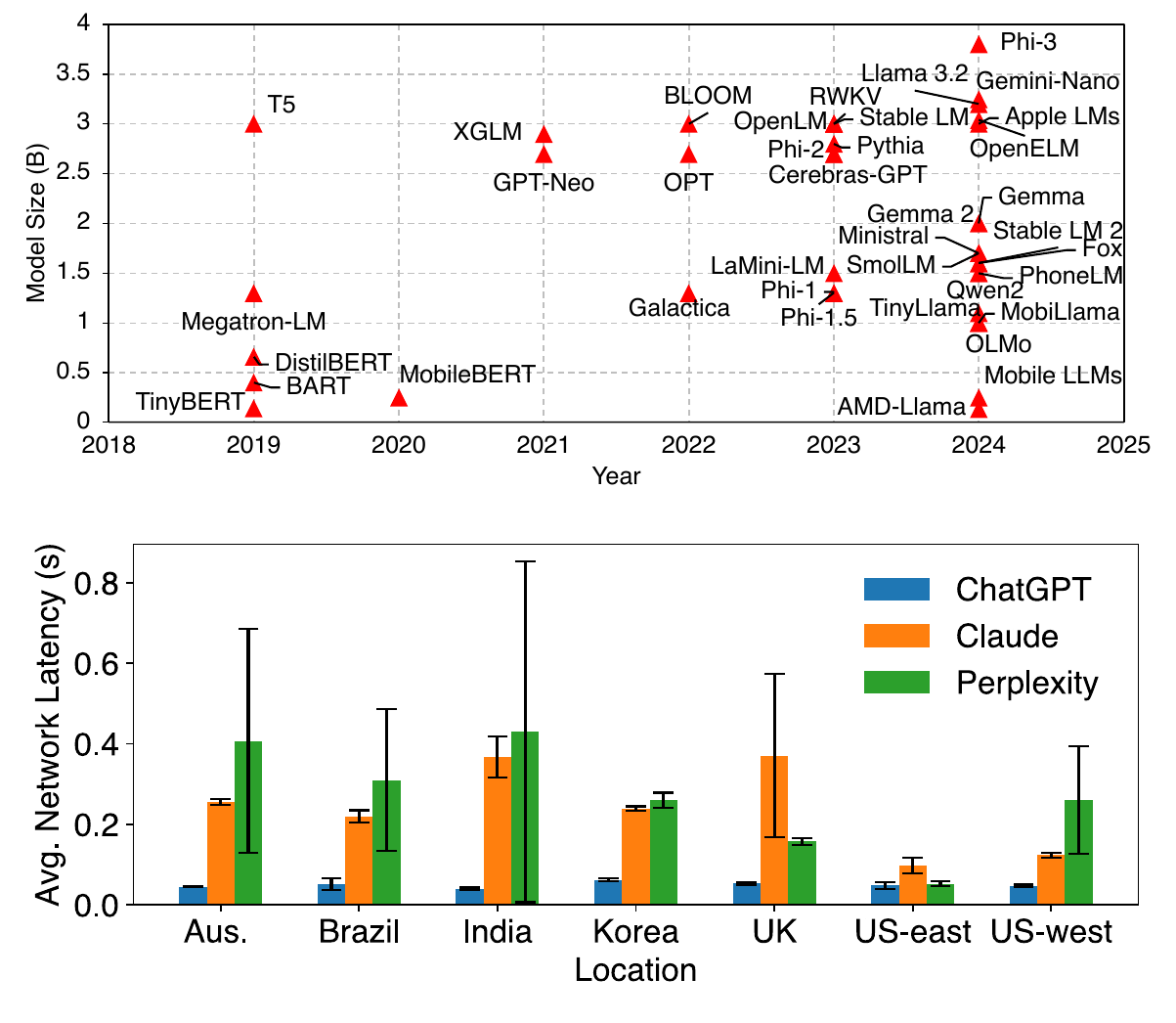}
    
    \caption{\textbf{Top}: Timeline of sub-4B SLMs designed for on-device deployments. Only the smallest version per model family is shown. \textbf{Bottom}: Average network latency to cloud-based LLM services (ChatGPT, Claude, Perplexity) from various global regions. Error bars indicate variability.}
    \vspace{-0.5cm}
    \label{fig:slms}
\end{figure}

Despite their remarkable performance and benefits, cloud-based LLM inference comes with their limitations. Firstly, the massive computational requirements of cloud-based LLMs result in high operational costs and significant $CO^2$ footprint due to the extensive energy consumption. For instance, training large-scale models like GPT-3 is estimated to consume the equivalent of the annual electricity usage of a U.S. household for 120 years \cite{adasci2024}, which is also equivalent to 1287 MWh of electricity, and resulted in carbon emissions of 502 metric tons~\cite{columbia2023co2}. From an LLM inference perspective, based on a recent study~\cite{luccioni2023estimating}, a similar size model BLOOM consumes on average 3.96 Wh per request, with energy usage scaling significantly as user demand increases.

Second, long latency is a common issue in cloud-based systems, as data must be transmitted over the network, processed remotely, and then returned, introducing delays that impact real-time applications. Figure~\ref{fig:slms}-bottom shows network latency measurement of various cloud-based LLM services. 
Popular LLM services, such as those provided by OpenAI, are hosted in centralized server locations (e.g., the US), which can impact the Quality of Experience (QoE) for users situated far from these hosting regions due to increased latency and reliability issues, with network latency ranging from around 48 milliseconds to hundreds of milliseconds depending on the service provider and user location. The Content Delivery Network (CDN) systems can help alleviate long latency delays by efficient network management, they cannot fully resolve the issue since the core processing of LLMs remains on remote infrastructure. 

Lastly, recent examples of prolonged service downtime \cite{openai_incident_2024} for popular LLM services highlight another critical limitation of centralized systems. This kind of outages can severely disrupt operations, especially for applications requiring high availability.

On the other hand, \textbf{edge-based SLM} inference aims to execute the model directly on-devices, reducing the dependency on the cloud. In response to such enormous cost of cloud-based LLMs, there has been advancements in model compression techniques, such as quantization \cite{lin2024awq-quantization}, pruning \cite{ma2023llmpruner}, and knowledge distillation \cite{gu2024llm-kd}, are helping to make edge AI a viable option for deploying SLMs on mobile \cite{laskaridis2024melting} and at the edge. For instance, Gemma2 is available in multiple versions, with the smaller model consisting of 2B parameters \cite{team2024gemma2}, enabling local laptops and even mobiles to run SLMs. Figure~\ref{fig:slms}-top shows recent trend of dramatic increased number of sub-4B language models over the past few years\cite{lu2024slm_survey}. Notably, these newer models are also becoming multi-modal, with capabilities extending beyond text processing to include image to text capabilities (Llama3.2-vision~\cite{llama-models}). This current shift reflects a broader industry trend toward developing versatile models that can handle a range of input types, enabling them to support complex application that require integrating information from multiple data sources-- capability especially necessary at the network edge.  
\setlength{\belowcaptionskip}{-5pt}
\begin{figure}[!t]
    \centering
    \includegraphics[width=.85\columnwidth]{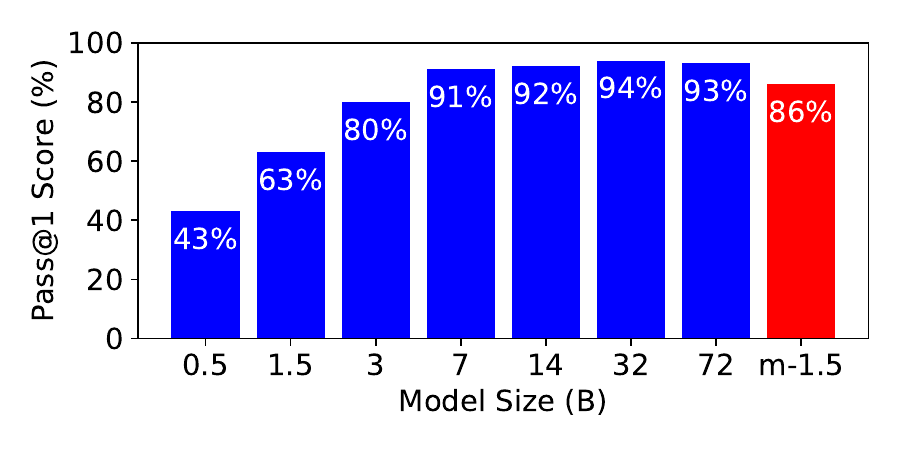}
    
    \caption{Pass@1 accuracy of LMs with varying parameter sizes on OpenAI gsm8k. Blue bars indicate general-purpose models; the red bar (m-1.5) shows a 1.5B fine-tuned model, highlighting accuracy gains from specialization.}
    \vspace{-0.3cm}
    \label{fig:accuracy}
\end{figure}

\setlength{\belowcaptionskip}{-5pt}
\begin{figure*}[!t]
    \vspace{-0.2cm}
    \centering
    \includegraphics[width=1\textwidth]{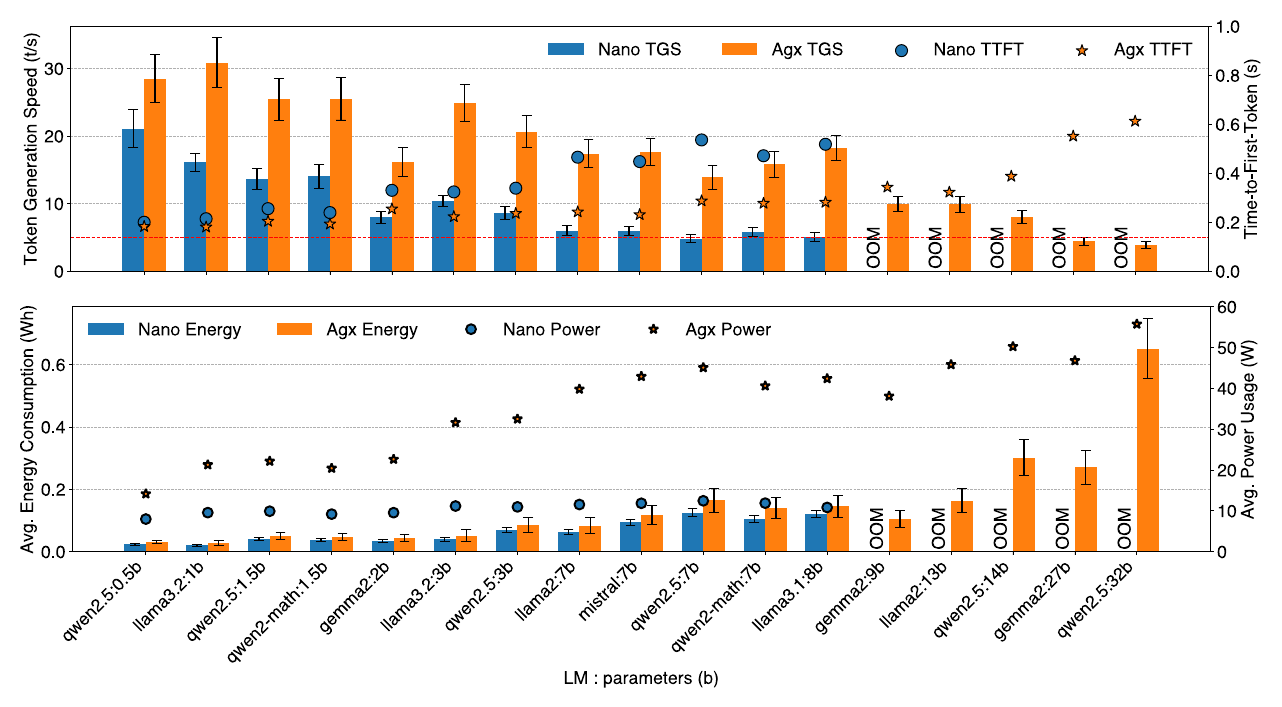}
    
    \caption{\textbf{Top}: Token generation speed (bars) and time-to-first-token (dots) on Jetson Nano (blue) and AGX (orange). The red dashed line marks human reading speed. \textbf{Bottom}: Power (dots) and energy per query (bars) on Nano and AGX.}
    \vspace{-0.2cm}
    \label{fig:tgs}
\end{figure*}
\section{Benchmarking SLMs on Edge Devices} 
\subsection{General vs. Specialized Models}
The challenge with deploying SLMs on mobile and edge devices lies in their limited computing capability, which directly restricts the loading of the weights to the memory. Since general purpose language models are inherently large, drastically reducing the model size leads to performance degradation. To evaluate the impact of the number of parameters on accuracy, we conducted an experiment using 100 queries sampled from OpenAI/GSM8K math test dataset, applying the length-stratified sampling technique to ensure a well-distributed query set. We then assessed the performance of the Qwen-2.5 language models, a versatile, general-purpose pretrained model. Qwen-2.5 comes in few variations, differentiated by the number of parameters: 0.5B, 1.5B, 3B, 7B, 14B, 32B, and 72B. We evaluated the performance of the model by comparing the model response to the correct answer.

In Figure~\ref{fig:accuracy}, the x-axis represents the number of parameters (in billions), while the y-axis illustrates the pass@1 accuracy, which indicates the percentage of queries answered correctly on the first try. As shown through blue bar graphs, SLMs with fewer parameters exhibit a significant drop in accuracy. Interestingly, accuracy improves with more parameters but plateaus around 91\%, where further increases yield only marginal gains.

A SLM pretrained on a specific domain may outperform a general purpose model when handling domain specific queries. In other words, rather than relying on an all-purpose model, a model trained or fine tuned on a specific domain may exhibit less knowledge across other areas or capability but will achieve greater accuracy within its fine-tuned field. For example, BERT-based models like HealthBERT for healthcare~\cite{kim2024health}, and FinBERT for finance~\cite{huang2023finbert}, are known to have less than 110M parameters fine tuned exclusively on their respective domains. These models excel in classification, identification, and prediction tasks. Similarly, Code Llama~\cite{roziere2023codellama7b}, DeepSeekMath~\cite{shao2024deepseekmath7b}, and BioMistral~\cite{labrak2024biomistral7b} all having 7B parameters, demonstrate exceptional capabilities in their respective domains: Code Llama excels in code generation, DeepSeekMath in mathematical reasoning, and BioMistral in medical question-answering tasks.

To assess this, we applied the same length-stratified sampling technique to the \textit{Qwen-2-math:1.5B} model, which shares the same architecture as Qwen-2.5 but is optimized for answering math questions. We conducted the same set of 100 queries, and the results, shown in the red bar graph labeled \textit{m-1.5} in Figure~\ref{fig:accuracy}, demonstrate that despite having only 1.5 billion parameters, the model achieves accuracy comparable to that of 7B general-purpose variant while requiring just 19.8\% of the model space. 

\subsection{On-device SLM inference} \label{sec:benchmark_tgs}
Computation capability of device in edge environment is also an important aspect of running various SLMs at the edge. 
A key performance metric in this context is token generation speed (TGS), time-to-first-token (TTFT), along with average power usage and energy consumption per query, all of which can vary significantly based on the hardware platform. TGS reflects the rate of token production during SLM inference which provides insights into throughput efficiency, while TTFT measures the latency for generating the first token which highlights the responsiveness of the model. To evaluate the performance of different models, we used two edge devices—Jetson AGX Orin and Jetson Nano Orin, commonly found in edge AI applications like drones and small robots, along with a smartphone—to run several SLMs. All experiments on the Jetson used its on-board GPU to execute SLM query inference.

Figure~\ref{fig:tgs}-top depicts the TGS and TTFT results for these two devices while running various SLMs. The x-axis represents the model’s parameter size, the left y-axis indicates TGS (in tokens per second), and the right y-axis indicates TTFT (in seconds). The orange indicates the Jetson Agx and blue indicates Jetson Nano. Notably, some models could not run on Jetson Nano due to memory limitations, and these cases are labeled as ‘OOM’ (out of memory) in the figure. The horizontal red line in Figure~\ref{fig:tgs}-top represents the average human words per minute reading speed, providing a baseline for decent TGS.  

While there are some discrepancies, models with fewer parameters typically generate tokens faster and require less time to generate the first token.  This is largely because they have fewer layers, which reduces the depth of sequential computations required for token generation (e.g., Qwen2.5:0.5b has 24 layers compared to Qwen2.5:32b with 64 layers). Additionally, embedding length plays a significant role in performance, as it directly affects the size of the matrices used in self-attention and feedforward layers. Larger embedding sizes result in more computational overhead and memory usage during inference, slowing down both token generation speed and time-to-first-token (e.g., Qwen2.5:0.5b has  896, while Qwen2.5:32b uses 5120). TTFT starts to increase as the number of model parameters approaches the device’s memory capacity (e.g., 7b for Nano and 27b for Agx), where memory constraints and computational overhead become significant bottlenecks.

Figure~\ref{fig:tgs}-bottom depicts average power consumption and energy consumption for processing a query. The x-axis represents the model’s parameter size, the left y-axis represent average energy consumption (in Wh), right y-axis represent average power consumption (in W). Jetson Agx consumes 1.7–3.91 times more power than the Nano; however, the Nano requires 1.49 times more time to generate a response. This time difference increases to 3.25 times for larger models, such as those with 7 billion parameters, where Nano approaches its performance limits. 
Despite this, on-device inference demonstrates significantly higher efficiency compared to cloud-based models. For comparison, an average energy consumption per query for cloud-based 176B parameter model like BLOOM, is estimated to be more than 3.96Wh \cite{luccioni2023estimating}. 

Results from Figure~\ref{fig:accuracy}, and Figure~\ref{fig:tgs} highlight that in edge computing environments, rather than relying solely on general-purpose shallow SLMs with limited performance, it becomes crucial to understand how to best deploy LMs across heterogeneous edge devices. Factors such as model specialization, device capacity, and workload characteristics have an important role in determining whether an edge deployment is sufficient, or whether cloud fallback becomes necessary. This motivates the need to define new metrics that can capture tradeoffs across multiple dimensions, which we explore next through the lens of energy, cost, inference latency, and response quality.
\section{Edge Efficiency, Cost and Quality Tradeoffs}
While metrics like service latency, response quality, and cost per query are well established for evaluating cloud-based LLMs, hey must be redefined for edge SLM agents to reflect variations in language model architectures as well as the unique characteristics and constraints of edge devices. On constrained edge devices, energy consumption differs significantly (Figure~\ref{fig:tgs}-bottom), and response quality varies by model architecture (Figure~\ref{fig:accuracy}). This implies that a combined metric would be necessary for capturing trade-offs between resource, cost, and response quality. This section introduces considerations for evaluation metric for SLM deployment at edge. 

From a system perspective, a holistic measurement of how the edge agent operates provides insights into its efficiency, hence we introduce an example metric \textbf{Performance-Cost Ratio (PCR)}, a metric that evaluates how effectively resources are used to deliver both quality and speed. To achieve this, we begin by defining a utility function ($U$) that combines quality ($Q$) and responsiveness ($R$), assigning a flexible weight \(\alpha\) to reflect their relative importance:
\vspace{-0.1cm}
\begin{equation}
\text{U} = \alpha \cdot Q + (1 - \alpha) \cdot R
\label{eq:utility_def}
\end{equation}

Here, \(Q\) represents the quality score (e.g., accuracy, BLEU, ROUGE, etc) and \(R\) indicates a responsiveness measure (e.g., TTFT or TGS, etc). The parameter \(\alpha\) controls the weight given to quality versus speed. 

Then, PCR can be defined as:
\begin{equation}
\text{PCR}_{\text{platform}} = \frac{\text{U}}{\text{CPR}}
\label{eq:pcr_def}
\end{equation}

Here, \textbf{Cost per Response (CPR)} represents the monetary cost of generating a single response. For cloud-based LLMs, this cost is directly determined by usage fees tied to API calls, often based on the number of tokens generated. 
    \begin{equation}
    \text{CPR}_{\text{cloud}} = \text{API usage in \textcent} 
    \label{eq:cpr_def_cloud}
    \end{equation}
    While edge operations do not incur direct service costs like cloud services, the device still consumes electricity, which can be translated into a monetary equivalent. Thus, for edge deployments, CPR can be calculated as:
    \begin{equation}
    \text{CPR}_{\text{edge}} = \text{Energy (kWh)} \times \text{Electricity Rate (\textcent/kWh)}
    \label{eq:cpr_def_edge}
    \end{equation}

\begin{table}[h!]
\centering
\caption{Comparison of platform metrics. CPR: Cost per request, Q: Query score (Accuracy), R: Responsiveness (Time to first token), PCR: Platform-cost ratio.}
\resizebox{\linewidth}{!}{%
\begin{tabular}{|l|l|l|l|l|l|}
\hline
Platform          & CPR (\textcent)  & Q & R & PCR     \\ \hline \hline
Cloud (GPT4)      & 1.65  & 0.97 & 0.71  & 0.51   \\ \hline
Edge (Agx, Qwen-2.5:7b) & 0.0041 & 0.91 & 0.28  & \textbf{145.12} \\ \hline
Edge (Nano, Qwen-2.5:3b) & 0.0017 & 0.80 & 0.33  & \textbf{332.35} \\ \hline
\end{tabular}%
}
\label{table:platform_comparison}
\end{table}

To demonstrate the practical application of these metrics, we compare a cloud-based LLM platform using GPT-4 with an edge-based SLM platform using Qwen-2.5:7b deployed on a Jetson Agx device and Qwen-2.5:3b deployed on a Jetson Nano device. On the cloud, an average input of 48 tokens (from GSM8K dataset) and output of 249 tokens for GPT-4 costs approximately 1.65 cents per response (Equation \ref{eq:cpr_def_cloud}). On the edge (Agx), energy consumption of 0.1646 Wh/query at \textcent25/kWh results in a CPR of 0.0041 cents/response (Equation \ref{eq:cpr_def_edge}), while on the edge Nano, 0.0687 Wh/query yields a CPR of 0.0017 cents/response. In terms of quality score  $Q$ , GPT-4 achieves 0.97 on GSM8K, compared to 0.91 for Qwen-2.5:7B and 0.80 for Qwen-2.5:3B. Responsiveness, measured as TTFT ( R ), is 0.71s for GPT-4, 0.28s for the Agx, and 0.33s for the Nano.

Finally, considering the PCR using $\alpha$ = 0.5 to weight quality and responsiveness equally, the combined utility scores (Equation \ref{eq:utility_def}) are 0.84 for GPT-4, 0.595 for Qwen-2.5:7B, and 0.565 for Qwen-2.5:3B. However, when normalized by their respective costs, the PCR values (Equation \ref{eq:pcr_def}) are 0.51 for the cloud (GPT-4), 145.12 for the edge Agx (Qwen-2.5:7B), and 332.25 for the Nano (Qwen-2.5:3B), making the edge platforms significantly efficient.

Despite the cloud’s higher quality and speed, the edge platform uses its resources extremely efficiently in this example, producing better combined performance per unit cost. This example demonstrates how these integrated metrics—CPR, and PCR—can illuminate nuanced trade-offs, enabling informed decisions about where and how to deploy language models for various use cases. 
\vspace{-0.2cm}
\section{Balancing Edge Limits and Cloud Re-directions} 
Although SLMs deployed at the edge (on-device) demonstrate strong performance under typical conditions, real-world deployments often encounter dynamic situations where resource scarcity, energy limitations, user preferences, or task complexity exceed the capabilities of on-device models. To manage these challenges, systems must not only balance local execution with selective offloading to cloud-based LLMs but also adopt mechanisms to optimize limited edge resources. Similar to how cloud-based LLM systems impose rate limits to manage resource usage and ensure fair access~\cite{ratelimits}, edge deployments must implement comparable strategies, particularly given the strict compute, memory, and power constraints typical of edge devices. These challenges are further amplified in heterogeneous environments, where devices vary significantly in capability, available energy budgets, and inference performance. 

In such contexts, \textit{edge rate-limiting strategies} serve a dual purpose: optimizing local resource usage while offloading excess demand to the cloud when necessary. To analyze the impact of these strategies on system behavior and cost, we developed a simulation framework that enables controlled experimentation under diverse conditions.

We focus on a \textit{sliding window approach}, which enforces rate limits over a rolling time window \cite{iqbal2020adaptive}. The window size is a critical parameter that governs system responsiveness and stability. A smaller window size enables the system to recover quickly from temporary overloads, making it particularly effective for handling bursty workloads (queries). In contrast, larger window sizes smooth out resource utilization over time but may lead to increased cloud redirection under sustained high-demand conditions.

Given the heterogeneity of edge devices, an additional challenge lies in how requests are assigned to devices for processing. We investigate the impact of three device selection methods that determine how incoming requests are distributed across available resources. 

\begin{itemize}
    \item \textit{Random method} requests are assigned uniformly to devices without consideration of their capacity.
    \item \textit{Weighted method} improves this by selecting device's probabilistically, favoring those with higher maximum capabilities. 
    \item \textit{Load-Aware method} dynamically directs requests to devices with the most available resources at a given time, mitigating overload and enhancing system efficiency.
\end{itemize}

We evaluate the sliding window strategy under realistic conditions with 30 users generating requests over an hour in one-minute steps. Workloads includes steady and bursty pattern, with bursts every 10 minutes with a higher probability of user activity. Token demands for each request are selected from a set of typical sizes (50, 100, 200, 300, and 500 tokens), weighted to favor smaller requests. The cloud cost for offloading requests is calculated as an average of OpenAI API pricing, considering both input and output token costs. The system comprises heterogeneous edge devices modeled on token generation performance (Figure \ref{fig:tgs}) with defined token capacities and parallel request limits. Each device is assigned a maximum token capacity and a limit on parallel requests. Table \ref{tbl:experimental} summarizes the experimental setup, including user activity, token demand distribution, and device capacities. Overall, our study considers four representative scenarios that combine workload types and device selection methods to understand their effect on request handling and cloud redirection. These scenarios include steady loads with random and capacity-weighted device selection, as well as bursty loads with random and load-aware selection methods. 

\begin{table}[h!]
\centering
  \caption{Experimental Simulation Setup}
  \label{tbl:experimental}
  \includegraphics[width=0.9\columnwidth]{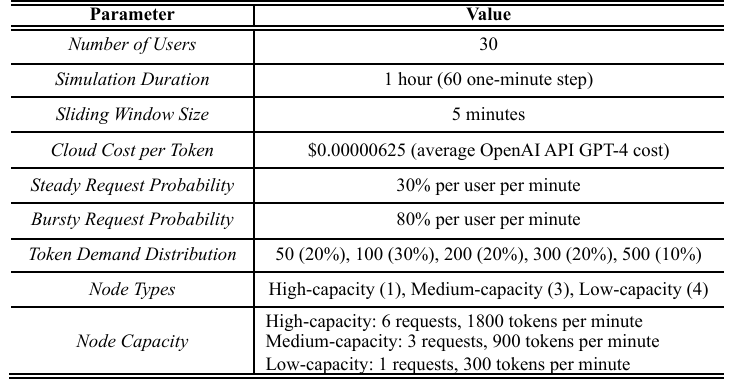}
  \vspace{-0.5cm}
\end{table}

\begin{figure*}[h]
    \centering
    \includegraphics[width=\textwidth]{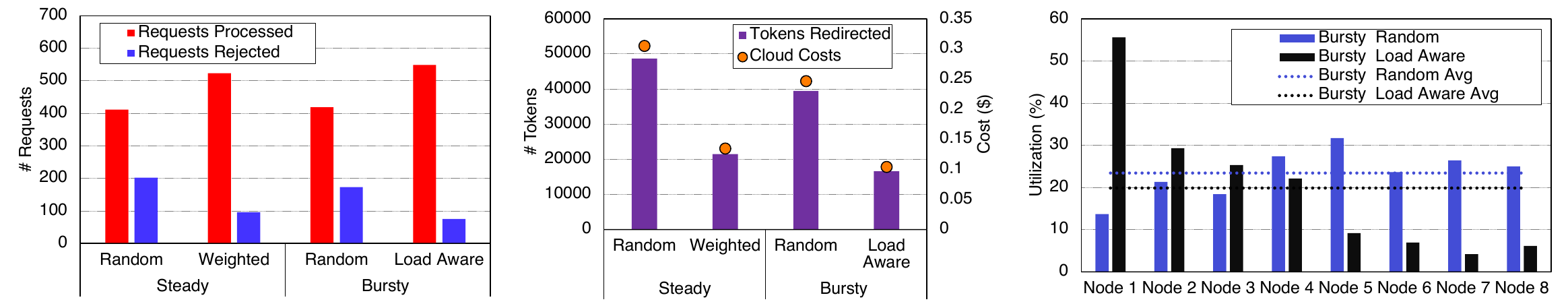}
    
    \caption{Performance comparison of device selection strategies under steady and bursty workloads with the sliding window rate-limiting approach. (Left) Number of requests processed and rejected. (Middle) Tokens redirected to the cloud and associated costs. (Right) Device utilization under bursty workloads, comparing the Random and Load-Aware strategies.}
    \label{fig:simulation}
\end{figure*}

Figure \ref{fig:simulation} illustrates the impact of device selection strategies (Random, Weighted, and Load-Aware) under steady and bursty workloads when combined with the sliding window rate-limiting approach. The first plot (left) shows that under steady workloads, the \textit{Weighted} processes the most requests but rejects fewer compared to \textit{Random}, highlighting its efficiency in leveraging high-capacity device. For bursty workloads, the \textit{Load-Aware} outperforms \textit{Random} by reducing the number of rejected requests, demonstrating its ability to dynamically distribute requests to underutilized devices' during high-demand periods. 

The second plot (middle) further quantifies the effect of device selection strategies on cloud redirection and costs. For steady workloads, \textit{Weighted} significantly reduces tokens redirected to the cloud and associated costs compared to \textit{Random}. In bursty workloads, the \textit{Load-Aware} leads to the lowest redirection and costs, highlighting its effectiveness in minimizing reliance on cloud resources during demand spikes. 

The third plot (right) focuses on device utilization under bursty workloads. The \textit{Random} strategy results in an imbalanced load across devices, with high variability in utilization, particularly underutilizing low-power devices. In contrast, the \textit{Load-Aware} achieves a more even distribution of load across the devices, reflected by its lower variability and closer alignment to the average utilization. This balanced load improves overall system efficiency and reduces cloud redirection.

These results reflect the specific workload characteristics and the types of devices used in this study and experimental setup, including heterogeneous edge devices with varying capacities. Further exploration of alternative configurations and combinations remains necessary to assess broader applicability
\section{Conclusion}
In this work, we assessed the limitations of cloud-based LLM inference and explored an edge-first perspective for deploying SLMs. Through empirical evaluation on off-the-shelf devices and distributed edge setups, we highlighted scenarios where SLMs offer competitive performance, and others where cloud fallback is still necessary. We introduced new metrics to capture tradeoffs across quality, latency, energy, and cost. Rather than promoting a one-size-fits-all solution, our results emphasize the importance of metric-driven decisions for efficient and adaptive LM inference across heterogeneous environments.
\vspace{-0.1cm}

\bibliographystyle{IEEEtran}
\bibliography{reference}

\end{document}